# Monitoring Extreme-scale Lustre Toolkit

Michael J. Brim
Computer Science & Mathematics Division
Oak Ridge National Laboratory
Oak Ridge, TN, U.S.A.
brimmj@ornl.gov

Joshua K. Lothian
Computer Science & Mathematics Division
Oak Ridge National Laboratory
Oak Ridge, TN, U.S.A.
lothian@ornl.gov

*Abstract*—We discuss the design and ongoing development of the Monitoring Extreme-scale Lustre Toolkit (MELT), a unified Lustre performance monitoring and analysis infrastructure that provides continuous, low-overhead summary information on the health and performance of Lustre, as well as on-demand, in-depth problem diagnosis and root-cause analysis. The MELT infrastructure leverages a distributed overlay network to enable monitoring of center-wide Lustre filesystems where clients are located across many network domains. We preview interactive command-line utilities that help administrators and users to observe Lustre performance at various levels of resolution, from individual servers or clients to whole filesystems, including job-level reporting. Finally, we discuss our future plans for automating the root-cause analysis of common Lustre performance problems.

## I. Introduction

Through its continued success in managing large-scale storage resources at numerous computing centers, including the Oak Ridge Leadership Computing Facility (OLCF), the San Diego Supercomping Center (SDSC), and the Texas Advanced Computing Center (TACC), Lustre has positioned itself as one of the few viable technologies for delivering the storage resources and performance that will be critical to the success of future extreme-scale computing systems. These future systems, whether in the form of exascale supercomputing systems for computational science or massive commodity clusters tasked with taming the Big Data deluge, are expected to push the Lustre storage architecture to its functional limits. Updating the Lustre software to achieve the desired levels of performance in the presence of unprecedented concurrency and new workloads such as data analytics is thus greatly important. An advanced Lustre performance monitoring and analysis framework will be key to helping Lustre architects, integrators, and administrators identify the source of performance and scalability problems as enhancements are developed, deployed, and evaluated.

In this paper we describe our ongoing efforts to design and develop the *Monitoring Extreme-scale Lustre Toolkit*, or MELT for short. MELT is designed as a unified Lustre performance monitoring and analysis infrastructure that provides continuous, low-overhead summary information on the health and performance of Lustre, as well as on-demand, in-depth problem diagnosis and root-cause analysis. It is this latter capability that is intended to separate MELT from currently available Lustre distributed monitoring solutions such as TACC's lltop/xltop [3][4] and the Lustre Monitoring Tool (LMT) [5] from Lawrence Livermore National Laboratory. These existing tools sample and optionally aggregate the Lustre-provided performance metrics [6] from the Linux proc file system on Lustre server and LNET router nodes. Unlike previous tools, MELT deploys monitoring agents across all Lustre nodes, including clients. Similar to TACC's tools, MELT interfaces with the job scheduling system to provide aggregate metrics on a per-job basis.

The rest of the paper is organized as follows. Section II reviews the design and architecture of MELT. Section III highlights the infrastructure and tool capabilities we are aiming to support. Section IV discusses the status of our implementation and our plans for evaluating the software on Spider2 [7], the production Lustre scratch file system serving all the computational resources at OLCF. Finally, Section V briefly relates our future plans for automating the root-cause analysis for common Lustre performance problems.

## II. Monitoring Extreme-scale Lustre Toolkit

### A. MELT Design Objectives

Our grand vision for MELT is to deliver a unified infrastructure for extreme-scale Lustre environments that simultaneously supports the persistent low-overhead performance and health monitoring needs of a computing center and the on-demand monitoring and performance problem diagnosis needs of system administrators and users. By allowing on-demand interactive tools to leverage the persistent distributed monitoring infrastructure, they can avoid suffering repeated distributed startup costs and utilize scalable distributed processing to accelerate the generation of useful information.

Unlike existing tools like LMT that monitor Lustre strictly using metric data available from server and router nodes, MELT also monitors Lustre client nodes in an attempt to identify client I/O behaviors that might negatively impact server performance as well as client-local resource contention.

This research was funded by the DoD-HPC program at Oak Ridge National Laboratory. This manuscript has been authored by UT-Battelle, LLC under Contract No. DE-AC05-00OR22725 with the U.S. Department of Energy. The United States Government retains and the publisher, by accepting the article for publication, acknowledges that the United States Government retains a non-exclusive, paid-up, irrevocable, world-wide license to publish or reproduce the published form of this manuscript, or allow others to do so, for United States Government purposes. The Department of Energy will provide public access to these results of federally sponsored research in accordance with the DOE Public Access Plan (http://energy.gov/downloads/doe-public-access-plan).

A key requirement for MELT is support for center-wide Lustre deployments, where one or more Lustre filesystems are shared by several computing systems. Generally, each of the computing systems resides in its own separate network domain, and thus the MELT infrastructure must be able to be deployed across all the compute and filesystem domains. Lustre itself handles cross-domain communication via LNET, but MELT must remain independent of LNET so as to not interfere with Lustre networking performance or protocols.

*B. MELT Architecture*

To meet our design objectives, the MELT architecture builds upon SNOflake, a general-purpose Scalable Network Overlay infrastructure being developed at ORNL. SNOflake's design targets next-generation exascale computing environments by evolving the concept of tree-based overlay networks (TBONs). TBONs such as MRNet [8] have been proven as a scalable and efficient communication and data processing architecture on leadership-class computing systems [2]. TBONs leverage the logarithmic properties of trees for scalable broadcast, multicast, and gather communication, and provide hierarchical distributed data processing within internal tree processes to support scalable data aggregations such as reductions and filtering. TBONs also provide inherent data redundancy for certain classes of data aggregations [1], which aids in fault tolerance and recovery.

As shown in Fig. 1, SNOflake extends the TBON architecture to a connected ring of TBONs, which enables the overlay network to span several resource domains. This capability enables SNOflake to serve as the basis for distributed applications, tools, system services, and middleware that require different classes of resources located within separate network domains. A SNOflake overlay places one or more TBON manager processes within each distinct network domain. Each TBON manager acts as the root of a tree spanning similar resources within the domain. The manager processes are assumed to be located on nodes with inter-domain communication capability. The manager processes are then connected within an inter-domain ring to enable communication between domains. A session root process serves as the top-level aggregator of data from all domains, and is the point of contact for persistent or transient client processes. Session clients create data streams with customizable data aggregation, and can request to spawn agent processes on any domain covered by the session overlay. Agents may also be started independently (e.g., during system initialization). Once started, agents attach to the session overlay and can subscribe to data streams created by a client, which enables the agents to send or receive data on the streams.

Three types of agents are used for MELT, based on the role that the hosting node plays in Lustre. Lustre client agents are launched on computational and interactive service nodes. Lustre server agents are used on OSS, MDS, and MGS nodes. Lustre router agents are placed on LNET router nodes. Each type of agent knows what specific Lustre performance metric information is available on its associated node and has a default sampling rate for each of the metrics. The default sampling rates are chosen to support our goals for low-overhead operation during continuous monitoring. To support higher metric resolution during on-demand, in-depth problem diagnosis, client tools can temporarily increase the sampling rate for one or more metrics. The agents sample the Lustre performance metrics at the current requested rate, and also periodically gather node-level performance information such as processor loads and memory utilization that can help in identifying node-local resource contention that negatively impacts Lustre performance.

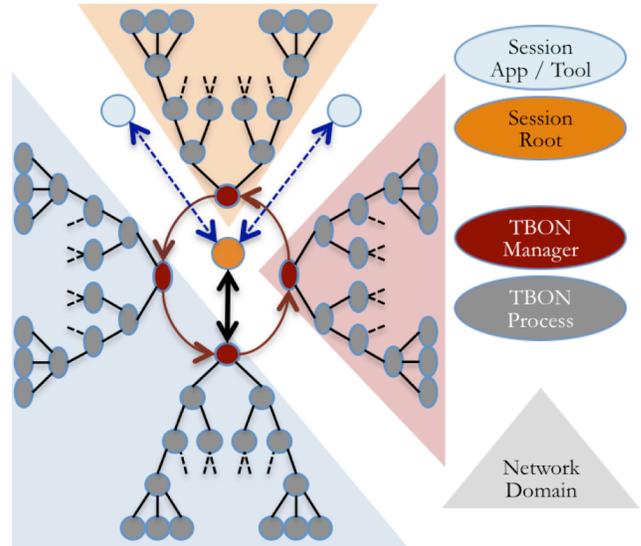

Fig. 1. **SNOflake Architecture.** A scalable cross-domain overlay network organized as a ring of tree-based overlay networks.

We envision MELT as a persistent service, so the SNOflake overlay and MELT agents are expected to be launched during system initialization. The overlay configuration is currently described in shared files that specify the available nodes and suggested topology for each domain, the node locations and ring topology for the TBON managers, and the node location of the session root. For MELT, it is expected that the session root process will reside on the MGS.

### III. MELT INFRASTRUCTURE & TOOLS

The MELT infrastructure consists of a SNOflake overlay network, including node-specific agents, deployed across all Lustre client, server (i.e., MGS, MDS, and OSS), and LNET router nodes.

After the MELT infrastructure has completed bootstrapping, a persistent monitoring daemon (`meltmon`) connects to the session root and creates data streams that use custom aggregations to summarize client, server, and router performance metrics. Once streams are published, they will exist and can be used by agents even if `meltmon` fails or disconnects. During periods where no session client is connected to the session root, SNOflake will simply buffer aggregated stream data (up to a configured maximum size) in anticipation of another client connecting and reading from the streams. The `meltmon` daemon is also responsible for interfacing with the job scheduling system(s) to periodically (e.g., once a minute) identify running jobs and their constituent nodes and multicast this information to the agents for use in job-level metric aggregations. For our initial implementation,

`meltmon` will simply write summarized metric data to a set of MELT performance logs. In future work, `meltmon` could be enhanced to push its data into existing infrastructure monitoring solutions such as NAGIOS, Ganglia, or Splunk, or to store it to a SQL database similar to LMT.

The MELT tools are interactive command-line utilities that attach to the MELT session. Once attached, tools generally follow one of a few common action patterns. The first pattern is to subscribe to the existing metric sampling streams created by `meltmon` and to display the summarized information. The remaining patterns require the tool to create new data streams with custom aggregation. The second pattern is to process the metric data (as collected by agents using the default sampling schedule) in new ways, such as the generation of histograms rather than the statistical summaries generated by default. The third pattern is to request an increased sampling rate for one or more metrics in order to obtain higher-resolution data, with the option to additionally process the data in new ways. The final common pattern is to request additional non-Lustre data from agents and to combine that data with Lustre metrics in a custom data aggregation stream.

The current design for the MELT tools relies on a single `melt` command-line utility with modes, metric classes, and various options. The projected usage of `melt` follows:

melt [*options*] *target mode classes* [*mode-opts*]

Table I shows a tentative set of targets, modes, and metric classes and gives a brief description of the resulting tool functionality. Each metric class will display all associated metrics by default. To select a subset of the class metrics, the names can be supplied using the mode option `metrics=name1,name2`. For all tools, we plan to provide output in both human- and computer-friendly (i.e., easily parsed) formats. Human consumption is assumed by default, while computer formats such as key-value pairs or syslog may be requested using the option `format={csv|kv|log}`.

The `status` mode periodically displays the selected metrics for the given metric classes. The `top` mode acts similar to `status`, but presents the results sorted by a key metric that may be specified. For the `status` and `top` modes, the delay between samples may be passed as a mode option. For the `top` modes, the value *k* indicating the number of requested entries may be passed as a mode option. All `status` modes also accept the metric class keyword `all` to indicate all available information should be displayed.

Various forms of grouping are supported using the `melt group={client|job|ost|server}` option. Available grouping methods depend upon the target and mode. For example, when viewing data from a particular OSS, results can be aggregated per-client, across all clients associated with each job, or across OSTs.

To highlight the utility of the MELT tools, we show how the tools could be used for a few use cases and provide corresponding simulated output. The first use case, shown in Fig. 2, is a Lustre administrator requesting a high-level overview of the I/O and metadata performance for all Lustre filesystems. In the second use case, shown in Fig. 3, the same

TABLE I. MELT COMMAND-LINE TOOL MODES

| TARGET | MODE | METRIC CLASSES | DESCRIPTION |
|---|---|---|---|
| fs | status top | io lock meta rpc | Aggregate or top-k I/O, locking, metadata operation, or RPC rates for one or all filesystems; grouped by server or job |
| job | status top | io meta | Aggregate or top-k I/O or metadata operation rates across all processes in a given job; optionally grouped by server |
| oss | status top | io lock rpc | Aggregate or top-k I/O, locking, or RPC service rates for a given server; optionally grouped by client, job, or OST |
| mds | status | lock meta | Aggregate locking or metadata operation rates for a given server; optionally grouped by client or job |
| mds | top | client op path | Top-k clients, metadata operation types, or paths accessed for a given filesystem |
| clnt | status top | io meta load rpc | Aggregate or top-k I/O, metadata operation, or RPC rates, or node loads for a given client; grouped by server or job |

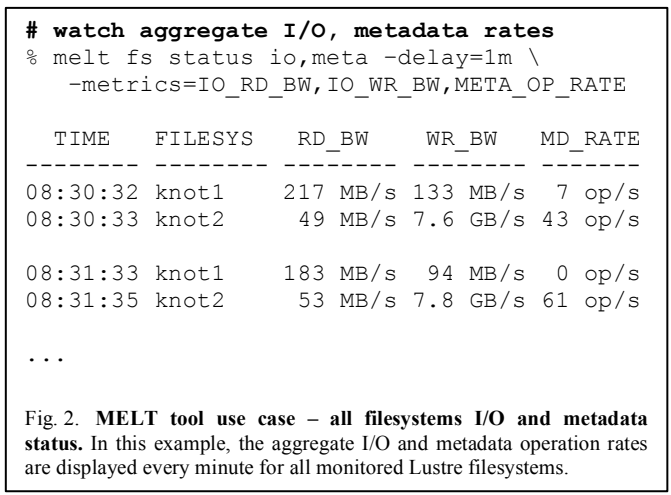

```
# watch aggregate I/O, metadata rates
% melt fs status io,meta -delay=1m \
  -metrics=IO_RD_BW,IO_WR_BW,META_OP_RATE

  TIME     FILESYS    RD_BW    WR_BW    MD_RATE
--------  --------  --------  --------  -------
08:30:32  knot1      217 MB/s  133 MB/s   7 op/s
08:30:33  knot2       49 MB/s  7.6 GB/s  43 op/s

08:31:33  knot1      183 MB/s   94 MB/s   0 op/s
08:31:35  knot2       53 MB/s  7.8 GB/s  61 op/s

...
```

Fig. 2. **MELT tool use case – all filesystems I/O and metadata status.** In this example, the aggregate I/O and metadata operation rates are displayed every minute for all monitored Lustre filesystems.

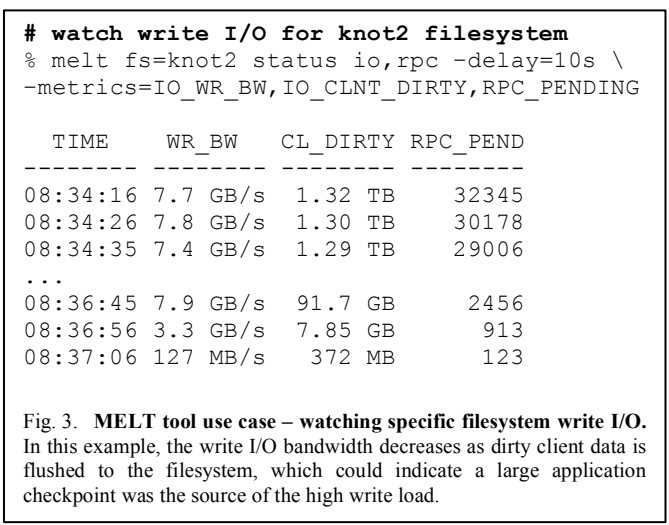

```
# watch write I/O for knot2 filesystem
% melt fs=knot2 status io,rpc -delay=10s \
 -metrics=IO_WR_BW,IO_CLNT_DIRTY,RPC_PENDING

  TIME     WR_BW    CL_DIRTY  RPC_PEND
--------  --------  --------  --------
08:34:16  7.7 GB/s   1.32 TB    32345
08:34:26  7.8 GB/s   1.30 TB    30178
08:34:35  7.4 GB/s   1.29 TB    29006
...
08:36:45  7.9 GB/s   91.7 GB     2456
08:36:56  3.3 GB/s   7.85 GB      913
08:37:06  127 MB/s    372 MB      123
```

Fig. 3. **MELT tool use case – watching specific filesystem write I/O.** In this example, the write I/O bandwidth decreases as dirty client data is flushed to the filesystem, which could indicate a large application checkpoint was the source of the high write load.

```
                    (a)  admininstrator logging job-level I/O every 5 minutes

% melt -group=job -format=log fs status io -delay=5m

Jan 15 11:22:33 skein melt[123]: job=tait.1111 IO_RD_BW=20M/s IO_WR_BW=476M/s \
    IO_CLNT_NUM=256 IO_CLNT_DIRTY=4.3G IO_CLNT_AVG_RD_SZ=776K IO_CLNT_AVG_WR_SZ=1M ...
Jan 15 11:22:33 skein melt[123]: job=tait.1113 IO_RD_BW=89M/s IO_WR_BW=21M/s \
    IO_CLNT_NUM=64 IO_CLNT_DIRTY=1.2G IO_CLNT_AVG_RD_SZ=507K IO_CLNT_AVG_WR_SZ=123K ...
Jan 15 11:22:33 skein melt[123]: job=tait.1114 IO_RD_BW=364M/s IO_WR_BW=28M/s \
    IO_CLNT_NUM=32 IO_CLNT_DIRTY=86M IO_CLNT_AVG_RD_SZ=1.4M IO_CLNT_AVG_WR_SZ=67K ...
...
Jan 15 11:27:37 skein melt[123]: job=tait.1113 IO_RD_BW=52M/s IO_WR_BW=156M/s \
    IO_CLNT_NUM=64 IO_CLNT_DIRTY=5.5G IO_CLNT_AVG_RD_SZ=27K IO_CLNT_AVG_WR_SZ=509M ...
Jan 15 11:27:37 skein melt[123]: job=tait.1114 IO_RD_BW=364M/s IO_WR_BW=28M/s \
    IO_CLNT_NUM=32 IO_CLNT_DIRTY=86M IO_CLNT_AVG_RD_SZ=1.4M IO_CLNT_AVG_WR_SZ=67K ...
Jan 15 11:27:37 skein melt[123]: job=tait.1117 IO_RD_BW=364M/s IO_WR_BW=0B/s \
    IO_CLNT_NUM=1024 IO_CLNT_DIRTY=0B IO_CLNT_AVG_RD_SZ=509M IO_CLNT_AVG_WR_SZ=0B ...
...

                    (b)  administrator monitoring top-5 jobs performing read I/O on knot2 filesystem

% melt -group=job fs=knot2 top io -topk=5 -topmetric=IO_RD_BW \
    -metrics=IO_RD_BW,IO_CLNT_AVG_RD_SZ,IO_CLNT_AVG_RD_TIME

    JOB         RD_BW     RD_SZ     RD_TIME
------------ --------- --------- ---------
conway.2789   12 GB/s   127 MB   63.9 ms
tait.4321    7.8 GB/s   156 MB   72.3 ms
euler.22397  7.2 GB/s   112 MB   64.5 ms
tait.4334    3.4 GB/s   354 MB    283 ms
euler.22388  780 MB/s  31.9 MB   54.7 ms

                    (c)  user watching job I/O and metadata operations

% melt job=tait.1234 status io,meta -delay=30s -metrics=IO_RD_BW,IO_WR_BW,META_OP_RATE

  TIME    RD_BW    WR_BW    MD_RATE
-------- -------- -------- --------
22:34:16 692 MB/s   0 B/s   64 op/s
22:34:46 417 MB/s  13 MB/s  33 op/s
22:35:15  71 MB/s  29 MB/s   6 op/s
...
```

Fig. 4. **MELT tool use case – various job-oriented I/O monitoring.** (a) administrator logging I/O rates for all jobs every 5 minutes, (b) administrator viewing an iotop-inspired display of top-5 job I/O rates every minute, and (c) user watching I/O rates for a specific job every 30 seconds

administrator may focus on watching a particular filesystem that is exhibiting higher than normal aggregate I/O load to help diagnose the cause. The final use case, shown in Fig. 4, presents three forms of job-oriented I/O monitoring.

IV. MELT STATUS & EVALUATION PLANS

A prototype of SNOflake has been developed and deployed on our testbed in the Extreme Scale Systems Center at ORNL. Development of the MELT infrastructure and tools is ongoing.

The testbed consists of a login node (skein), three small compute clusters (16-node tait, 2-node conway, 32-node euler) and a shared Lustre filesystem. The testbed Lustre setup includes eight OSS nodes connected to a SAN, six LNET router nodes, and two MGS/MDS nodes in a fail-over pair. The LNET routers bridge between two Mellanox FDR Infiniband switches. One switch has connections to the compute clusters, and the other has connections to the server nodes.

We configure the SNOflake overlay as five domains: one for each compute cluster, one for the LNET routers, and one for the OSS nodes. A small TBON is deployed on the nodes of each domain, the TBON managers are connected in a ring, and the session root is placed on the login node. All overlay interprocess communication is via TCP sockets, which relies on IPoIB as necessary.

Our evaluation plan is to deploy the MELT infrastructure and tools on the testbed, where we will refine their functionality, usability, and performance based on feedback from the OLCF Lustre administrators. In late spring of 2015, we are planning to work with the OLCF staff on a large-scale

MELT testshot spanning the Spider2 filesystems and the Titan (Cray XK7) and Rhea (Dell cluster) compute systems. During this testshot, we plan to measure the processing, memory, and network utilization overheads of MELT's agents on clients and servers while running `meltmon` in its various configurations for continuous monitoring. The results of the testshot will be used to further refine default sampling rates and the MELT overlay topology.

## V. MELT Root-cause Analysis

A common scenario on existing large-scale Lustre deployments is a user reporting "Lustre is slow" after observing an application phase containing file I/O that runs noticeably longer in duration than some prior execution. In response, a Lustre administrator will likely perform a cursory health check on filesystems and a quick inspection of the Lustre server node resource consumption levels. If nothing obvious stands out as a problem, the user is typically greeted by a somewhat canned response to the effect that one or more of the following situations may be in play:

- Lustre is a shared resource and your job is seeing contention from other jobs at the MDS or OSSs.
- The compute and/or storage networks are a shared resource and your job is seeing congestion, possibly due to bad luck in the job's node placement.
- Are you sure it is the I/O that is slow?

Of course, none of these answers really help users or ease their concerns, but from the administrator's viewpoint it is more polite than "stop bugging us".

To decrease the annoyance levels of both administrators and users, it would be beneficial to provide automatic root-cause analysis of performance problems. Although the MELT tools provide a great deal of utility, identifying the root-cause of a particular performance problem still requires some level of expertise and past experience in diagnosing previous problems. Our eventual goal for MELT is to embed such expertise within the toolkit, such that an administrator could simply invoke the `oracle` mode on a given filesystem, client, server, or job target, and the tool would automatically investigate potential problems and report back one or more identified causes. The administrator could then use this information to generate an informed response to users reporting problems. Depending on the level of performance overhead induced by the `oracle` mode, the root-cause analysis functionality could even be provided directly as user-oriented tools to head off problem reports before they are sent.

## Acknowledgments

The authors would like to thank Brad Settlemyer for his efforts to conceptualize and direct the initial phase of this work. We thank the OLCF's Jason Hill and Don Maxwell for their suggestions and insights into desired Lustre performance monitoring capabilities.

This work used resources of the Extreme Scale Systems Center, supported by the Department of Defense, and the Oak Ridge Leadership Computing Facility, supported by the Office of Science of the Department of Energy, at Oak Ridge National Laboratory.